\documentclass[useAMS,usenatbib,usegraphicx]{mn2e}

\title[A Pixel Approach Towards Understanding Galaxies]{Structure Through Colour:A Pixel Approach Towards Understanding Galaxies}
\author[M. M. Lanyon-Foster, C. J. Conselice, M. R. Merrifield]{M. M. Lanyon-Foster$^{1}$\thanks{E-mail:
ppxmml@nottingham.ac.uk}, C. J. Conselice$^{1}$, M. R. Merrifield$^{1}$ \\
$^{1}$University of Nottingham, School of Physics \& Astronomy, Nottingham, NG7 2RD UK}

\begin{document}

\date{Accepted ; Received ; in original form}

\pagerange{\pageref{firstpage}--\pageref{lastpage}} \pubyear{2007}

\maketitle

\label{firstpage}

\begin{abstract}
We present a study of pixel Colour Magnitude Diagrams (pCMDs) for a sample of
69 nearby galaxies chosen to span a wide range of Hubble types. Our goal is to
determine how useful a pixel approach is for studying galaxies according to
their stellar light distributions and content. The galaxy images were analysed
on a pixel-by-pixel basis to reveal the structure of the individual pCMDs. We
find that the average surface brightness (or projected mass density) in each
pixel varies according to galaxy type. Early-type galaxies exihibit a clear
``prime sequence'' and some pCMDs of face-on spirals reveal ``inverse-L''
structures. We find that the colour dispersion at a given magnitude is found
to be approximately constant in early-type galaxies but this quantity varies
in the mid and late-types. We investigate individual galaxies and find that the
pCMDs can be used to pick out morphological features. We discuss the discovery
of ``Red Hooks'' in the pCMDs of six early-type galaxies and two spirals and
postulate their origins. We develop  quantitative methods to characterise the
pCMDs, including measures of the blue-to-red light ratio and  colour
distributions of each galaxy and we organise these by morphological type. We
compare the  colours of the pixels in each galaxy with the stellar population
models of Bruzual \& Charlot (2003) to calculate star formation histories for
each galaxy type and compare these to the stellar mass within each pixel. Maps
of pixel stellar mass and mass-to-light ratio are compared to galaxy
images. We apply the  pCMD technique  to three galaxies in the Hubble Ultra
Deep Field to test the usefulness of the analysis at high redshift. We propose
that these results can be used as part of a new system of automated
classification of galaxies that can be applied at high redshift.
\end{abstract}


\section{Introduction}

For as long as we have known about galaxies, astronomers have striven to
classify them in a physically meaningful way. The long-standing method devised
by Hubble (1926, 1936), with later revisions by de Vaucouleurs (1959) and
Sandage (1961), has, until recently, been apt for this purpose. The Hubble
sequence of morphological types is, on average, accurate for grouping together
galaxies by structure and tells us about the typical distribution of stellar
ages within a galaxy, as well as other properties (Roberts \& Haynes
1994). However, classification is not an exercise to be carried out purely for
its own sake; by classifying galaxies we hope to naturally obtain a picture of
galaxy evolution. Furthermore, with the rapidly growing amount of data
obtained at high redshifts, eyeball methods of classification are no longer
viable, and the Hubble sequence breaks down for describing the galaxy
population (e.g., Conselice, Blackburne, Papovich 2005, Conselice 2006b).

High redshift galaxies are also not adequately described by the Hubble
sequence, as the majority of faint and/or distant galaxies are peculiar
(Driver, Windhorst \& Griffiths 1995, Driver et al. 1998, Abraham et al. 1996;
Conselice et al. 2003, 2005, Lotz et al. 2006). At increasingly high
redshifts, fewer and fewer galaxies appear to resemble nearby systems. Galaxy
mergers, for example, become much more common at high-z (e.g., Conselice et
al. 2003; Lin et al. 2004; Cassata et al. 2005; Bundy et al. 2005; Conselice
2006; Bridge et al. 2007), and do not easily fit into any category of existing
classification systems. Clearly, a new method of classification is needed that
is physically meaningful and can naturally encompass and arrange all galaxy
types, independent of redshift, that can be applied automatically to data with
little or no human intervention.

Although the basic Hubble scheme has been revised and extended over the years
to include increasingly complex morphological structures within galaxies
(e.g., de Vaucouleurs et al. 1959; van den Bergh 1960, 1995), these features
are rarely seen at high redshift and do not account for the peculiar galaxies
seen. A classification based on quantitative, objective indices are more
suitable for classifying and quantifying these distant resolved galaxies.

The first attempt to develop such a scheme for nearby galaxies was made by
Morgan (1958, 1959), who proposed a system based on a measure of the central
concentration of light. This work has more recently been built upon by e.g.,
Bershady, Jangren and Conselice (2000), Conselice, Bershady and Jangren (2000)
and Conselice (2003) who have extended this idea by developing the
concentration, asymmetry and clumpiness (CAS) parameters. These parameters
allow an extension of classification to objects over a wide range of redshifts
and  morphologies.  However, these non-parametric indices do not utitlise all
the information presented in the resolved images of galaxies.

As such, we describe a new method of analysing the structures of galaxies on a
pixel-by-pixel basis, based on  similar method first used by Bothun (1986),
Abraham et al. (1999) and Eskridge et al. (2003).   Our approach is very
general and we aim to determine how much information a pixel approach towards
understanding galaxies reveals.  We use photometric data to construct
individual Colour Magnitude Diagrams of nearby galaxies such that each point
corresponds to one pixel of a galaxy image. These pixel Colour Magnitude
Diagrams (pCMDs) are used to examine the stellar populations and structure of
the galaxies and we discuss ways this technique can be applied at high
redshift.

We find that pCMDs provide information about galaxy morphology, in particular
early-type galaxy pCMDs are very different to those for late-type
galaxies. The pCMDs can be used to pick out individual features and this has
led to the discovery of ``Red Hooks'' in six early-type galaxies. We find that
quantitative parameters can be obtained from pCMDs, which can potentially be
used as measures of classification. Comparisons to stellar population models
can also be made on a pixel-by-pixel basis, presenting a new perspective on
galaxy structure and evolution. We show that pixel-by-pixel analysis can be
applied to high redshift galaxies and this is briefly tested on three Hubble
Deep Field galaxies.

The pCMD method is explained and the sample is defined in \S 2. The resulting pCMDs
are presented and the main trends found are detailed in \S 3, and
individual cases are examined. A quantitative analysis of pixel light
and colour is made in \S 4 and comparisons to stellar population models are
made in \S 5. An application of the pCMD method is made to 3 UDF galaxies in
\S 6 and our conclusions are discussed in \S 7.

\section{The Dataset and Method}
It is essential in any work on galaxy classification that a wide range of Hubble
types are considered. It is primarily for this reason that we select the
Frei et al. (1996) sample of nearby galaxies for this study. This is a digital
multiband sample of 113 local galaxies, made up of CCD images taken from
observations at the Lowell and Palomar Observatories. The galaxies in this
sample are all bright and well resolved, which is necessary for a structural
study such as this.

Images obtained at the Lowell observatory are available in B$_J$ (450nm) and R
(650nm) bands whereas the Palomar galaxies are obtained in g, r, and i bands
of the Thuan-Gunn photometric system. Our sample is refined to the 82 Lowell
images so that the colours of the galaxies can be accurately compared. We
remove 13 of the 82 galaxies from the sample due to low
signal-to-noise. Figure 1 illustrates the distribution of galaxy types in the
sample.

\begin{figure}
\includegraphics[width=84mm]{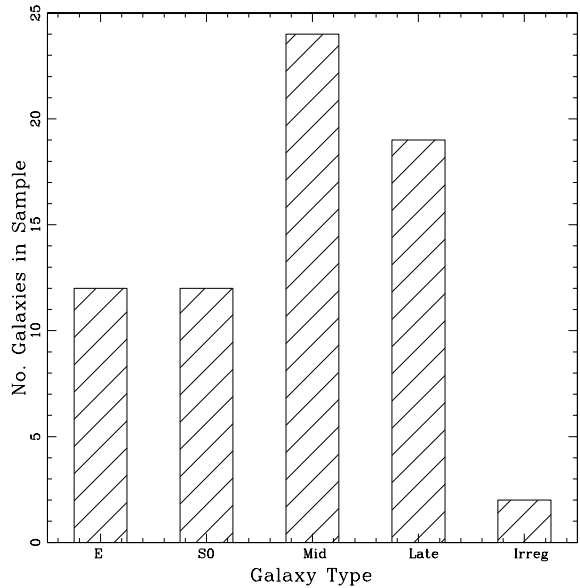}
\caption{The distribution of galaxy types in the sample.}
\end{figure}

\begin{figure*}
\centering
\includegraphics[angle=0, width=170mm, height=120mm]{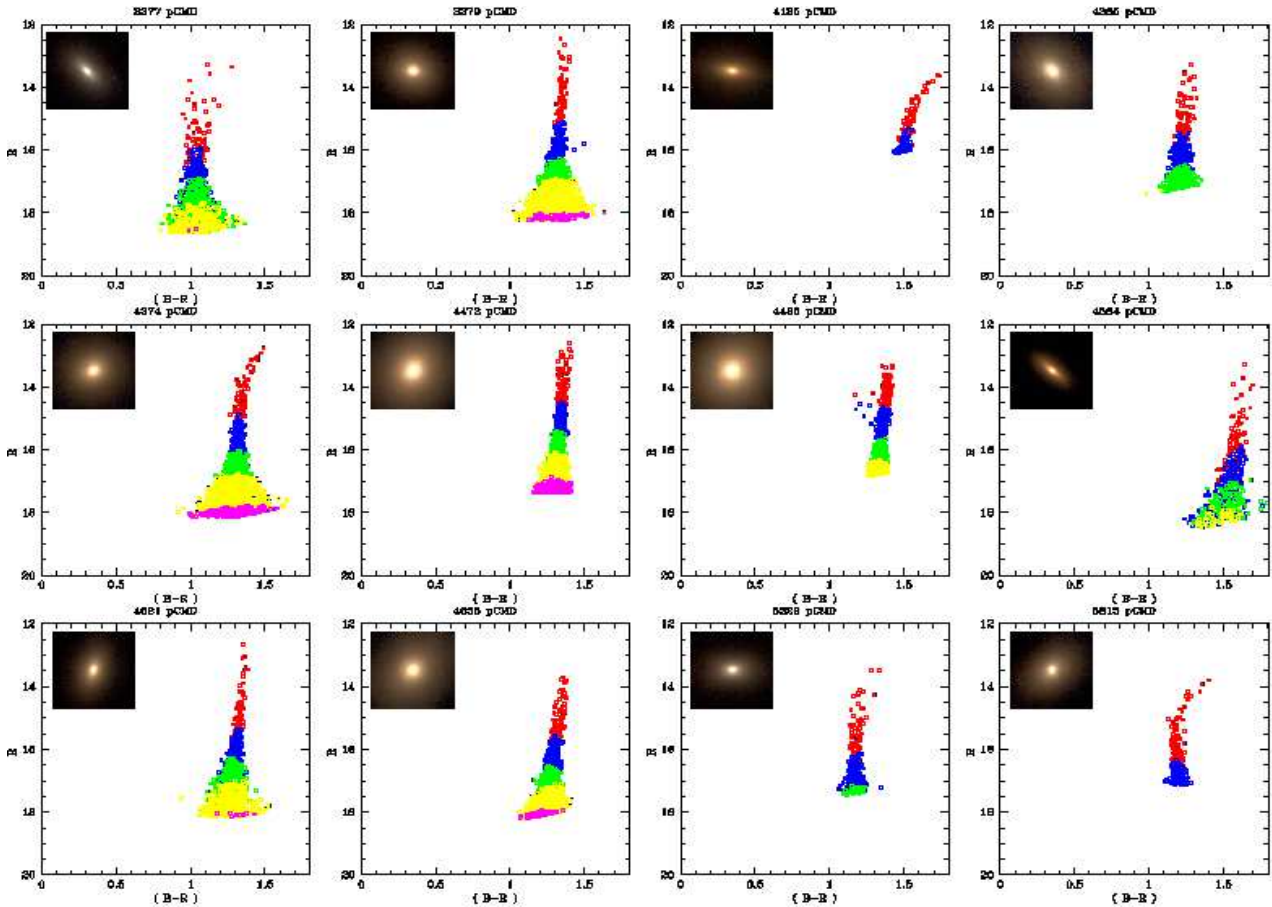}
\caption{The pCMDs of the elliptical galaxies in the sample. Each galaxy is
   displayed in its own pCMD and are colour coded by radius. Radius limits
   were set at 5, 10, 15, 25, 35 and 50 pixels from the centre, corresponding
   to the red, blue, green, yellow, magenta, cyan and black data points respectively.}
\end{figure*}
\newpage

\begin{figure*}
\centering
\includegraphics[angle=0, width=170mm, height=120mm]{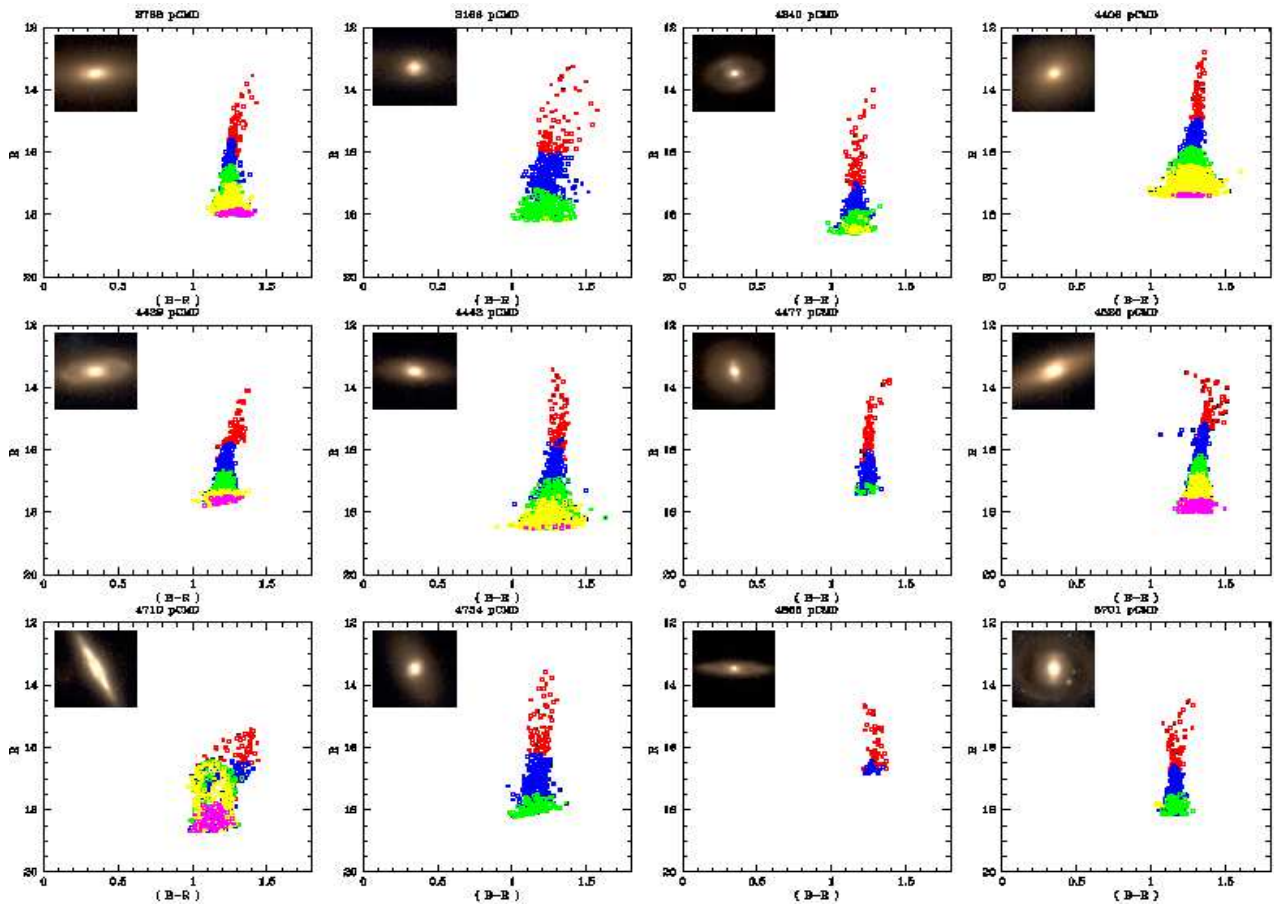}
\caption{The pCMDs of the S0 galaxies in the sample. The colour coding of the pixels is 
the same as in Figure 2.}
\end{figure*}
\newpage

\begin{figure*}
\centering
\includegraphics[angle=0, width=170mm, height=203mm]{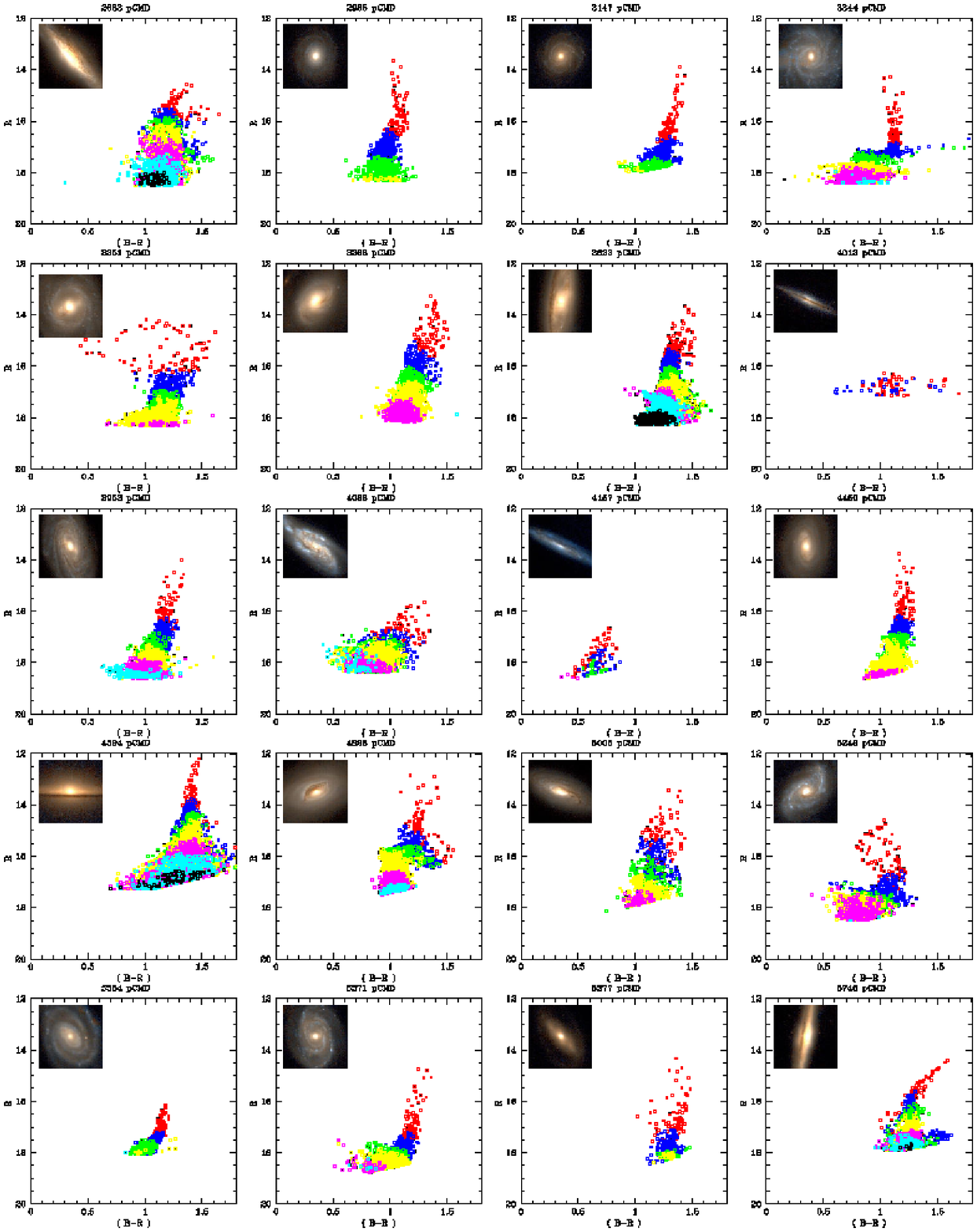}
\caption{The pCMDs of the mid-type spiral galaxies in the sample. The colour coding is the 
same as in Figure 2.}
\end{figure*}
\newpage

\begin{figure*}
\centering
\includegraphics[angle=0, width=170mm, height=40mm]{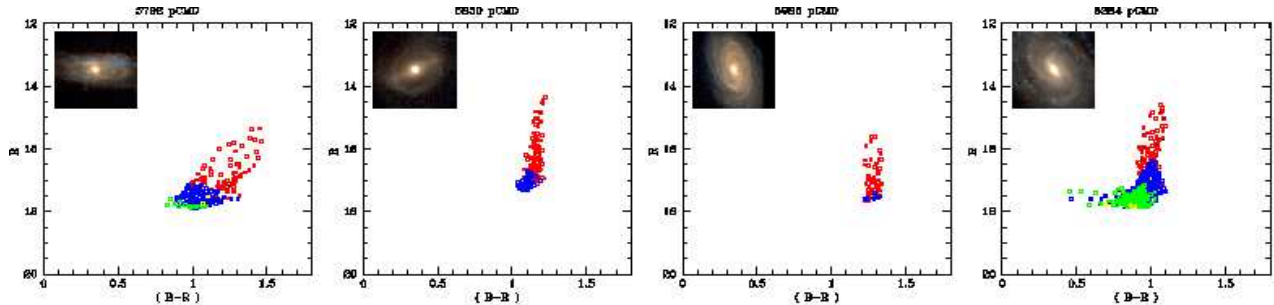}
\caption{The pCMDs of the mid-type spiral galaxies in the sample continued. }
\end{figure*}
\newpage

\begin{figure*}
\centering
\includegraphics[angle=0, width=170mm, height=204mm]{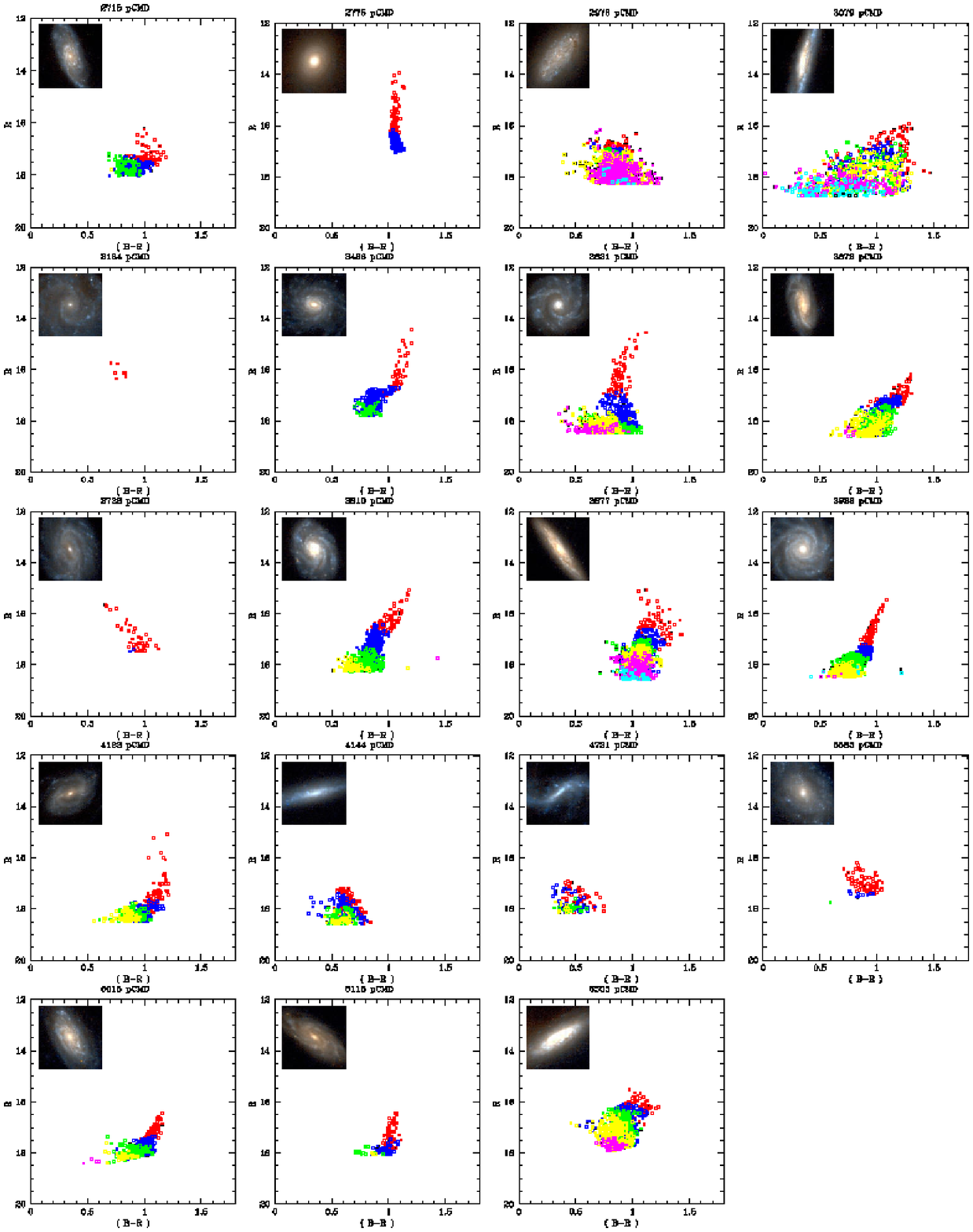}
\caption{The pCMDs of the late type spiral galaxies in the sample. The colour coding is the 
same as in Figure 2. }
\end{figure*}
\newpage

\begin{figure*}
\centering
\includegraphics[angle=0, width=80mm, height=40mm]{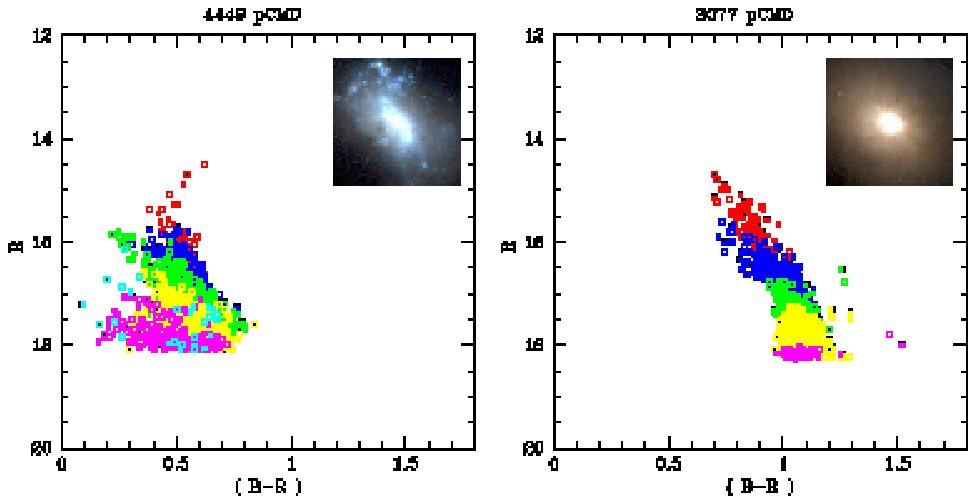}
\caption{The pCMD of the irregulars in the sample, NGC 4449 and NGC 3077. The colour coding 
is the same as in Figure 2.}
\end{figure*}
\newpage

The pCMDs are computed through the manipulation of the B$_J$ and R band galaxy
images using the Image Reduction and Analysis Facility (IRAF). The background
sky value is subtracted from all images. The centres of the images are found
and the R band images shifted such that the pixels are aligned with those
of the corresponding B$_J$ band image. Pixel alignment is crucial in this
method and we are able to match our images to an accuracy of $\pm$
10$^{-5}$ pixels.

The original pixel scale of the images in our subset of the Frei sample is
1.35 arcsec per pixel. The typical seeing for these galaxies is between two
and four arcseconds. In order to obtain a pixel scale that is at least on the
order of the PSF, we bin the the images by a factor of two. This ensures that
each binned pixel is statistically independent of the surrounding pixels and
improves the signal-to-noise.

Corrections are made for galactic extinction using values obtained from the
NASA/IPAC Extragalactic Database (NED) and a cut is made at a signal-to-noise
ratio of two. Fluxes are converted from counts per pixel to apparent magnitude
per square arcsecond by considering the pixel scale. Absolute magnitudes
are calculated using distances to the galaxies, which are taken from Conselice
et al. (2000). The (B$_J$ - R) colour index is calculated from each pixel in
the image and values for each galaxy are plotted on a separate pCMD.

Each galaxy is plotted on its own Colour Magnitude Diagram (CMD), with each
point on the CMD corresponding to one pixel of the galaxy image. The points
are colour-coded according to distance from the centre of the galaxy. The red
data points are the most central, falling within a radius of 5 pixels from the
centre of the galaxy. Radius limits are then set at 10, 15, 25, 35 and 50
pixels from the centre, corresponding to the blue, green, yellow, magenta,
cyan and black data points respectively.

The pixel Colour Magnitude Diagrams (pCMDs) represent a method of
illustrating the structure of a galaxy through its CMD. Indeed, we have
found that the shape of a galaxy's pCMD is strongly correlated to its
morphology and Hubble type. This will be explained in detail in Section 3.

We construct pCMDs for 69 galaxies in total, comprising of: 12 elliptical
galaxies (Figure 2), 12 S0s (Figure 3), 24 mid-types (Figures 4 \& 5), 19
late-types (Figure 6) and 2 irregular galaxies (Figure 7). Throughout this
paper we define early-type galaxies as both elliptical and lenticular,
mid-type galaxies as those on the sequence Sa-Sb(c) and late-types as those on
the sequence Sc-Sd(m). The classifications of Conselice et al. (2000) are used
throughout.

\section{Main Trends}

The pCMDs (Figures 2-7), despite appearing to vary widely, tend to have similar features
when sorted by galaxy type. In this section we will discuss the
various forms of the pCMDs and the trends found across the Hubble sequence.

\subsection{Early-Type Galaxies}

\subsubsection{Elliptical Galaxies}
The pCMDs of the ellipticals (Fig. 2) all have very similar qualities. They
all have pixel colours of (B - R) $\geq 1$ and display a narrow ``prime
sequence'' between colour and magnitude. We define this prime sequence as a
narrow trend between pixel colour and magnitude such that pixels that are
brighter are also redder.  The spread in colour of the ellipticals tends to be
small throughout the radius of the galaxy, the average colour dispersion at a
fixed magnitude being $0.029\pm0.014$ (see \S4.3). This is expected for the
ellipticals and is explained by their structure and homogeneous age. They are
old systems with mature stellar populations and little or no current star
formation (e.g. Bower et al. 1998).

\subsubsection{S0s}
The S0 sample (Fig. 3) yielded pCMDs with properties resembling those of the
ellipticals but with some subtle differences. The spread of colour of the
pixels throughout the radii of the S0s appears slightly larger than that of
the ellipticals in most cases, the average colour dispersion at a fixed
magnitude being $0.036\pm0.019$ (see \S4.3). This is likely to be accounted
for by considering that S0 galaxies tend to be dustier systems than elliptical
galaxies and such a spread in colour of pixels located at the same radius in
the galaxy appears to be strongly influenced by the presence of dust in the
galaxy (see \S3.2.1 \& \S3.3.3).

S0s are also well known to have bluer disks than bulges (e.g., Bothun \& Gregg
1990). This stellar population/metallicity effect could also potentially
account for the larger colour disperion at a fixed magnitude in the S0 pCMDs.

This phenonmenon could be of use in separating Es and S0s, a
notoriously tricky feat to accomplish by eye alone. It is possible that
S0s with a tight prime sequence may in fact be misclassified ellipticals
(e.g. NGC 4406, Fig. 3) and vice versa for those ellipticals with a large
uniform spread (e.g. NGC 3377 \& NGC 4564, Fig. 1). However, this assertion
will need to be tested on a larger sample of early-type galaxies before it can
be made with any certainty.

\subsection{Mid to Late-Type Galaxies}
The mid and late-type galaxy (Sa-Sd) pCMDs differ from the early-types. There are
general trends amongst these classes, such that the average position on the pCMD
is bluer than the early-types and has a larger colour gradient. The pixel colour
coding shows that the inner pixels of these galaxies are the reddest, with (B -
R) decreasing with increasing radius. The spread in colour is largest in the
outer parts of the galaxy. There is also a larger spread in the inner regions
of the spirals than for the early-types. Out to a five pixel radius
(red points in Figures 2-7) the average colour spread for the spirals is
$0.38\pm0.20$ magnitudes. In the same region in the early-types the average
spread is $0.21\pm0.10$ magnitudes.

Mid and late-type galaxy pCMDs are also not as uniform as those of the early-type
galaxies (Figs. 4-6). There are, however, some striking correlations between the
shape of the pCMD and the physical properties of the galaxies.

\subsubsection{Dusty and High Inclination Spirals}
pCMDs with a large spread in colour throughout the radius of the galaxy are
found to correspond to systems with a high inclination angle and dust
content. NGC 3079 (see Fig. 6) is a typical example of how much effect the
inclination of a galaxy can have on its pCMD. Star formation can be seen in
the outer edge of the disk, where the pixels are bluest. There is a colour
gradient similar to other mid and late-types but there is also more spread in
colour at all radii. That is, the prime sequence is not well defined for these galaxies.

There is a definite correlation between dust features in
the galaxy image and colour spread in the pixels (see \S3.3.3). It is evident
that dust can block light altogether if it is dense enough. In the pCMD of NGC
4013 (Fig. 4) the brightest light in the bulge of the spiral is absorbed by
the prominent dust lane that runs the length of the spiral and is most
noticeable due to the very high inclination of the galaxy.

\subsubsection{``Inverse-L'' Shapes}
Some pCMDs are found to have abrupt changes  in colour rather than a
continuous gradient. Such ``inverse-L'' shapes are found to correspond to
face-on spirals. The inner pixels appear very similar to typical elliptical
pCMDs with a prime sequence, while the outer pixels suddenly become much
bluer. Such behaviour is naturally explained by an old, red bulge surrounded
by blue, star-forming spiral arms. NGC 3344 (Fig. 4) clearly exhibits this
structure.\footnote{The odd branch of pixels that can be seen in the pCMD of NGC 3344, which
appears much redder than others at the same radius, correlates with a
foreground star that was removed in the B band image but not in the R
band. This demonstrates that unusual features in the pCMD do correspond with a
real feature, in this case a data reduction oversight. We discuss in \S 3.3 some
examples where physical structures within the galaxy have been detected in the
pCMDs.}

\subsection{Individual Cases}
Unusual features are picked out in some of the pCMDs. For example, in some
cases pixels that deviate from the main trend are found to correlate with
physical features of the galaxy. We discuss here some of the unusual pCMD
features that are discovered and explain them in terms of the structural and
physical properties of the galaxies.

\subsubsection{Central Red Pixels - The ``Red Hooks''}

We find that in the very central regions of 8 galaxies the colour gradient of the
pixels suddenly become much redder with decreasing magnitude than the outer
regions of the galaxy. These red pixels are, on average, within a radius of
2.5 pixels, or approximately 300pc, from the centre of the galaxy. These
galaxies are NGC 4125, NGC 4374, NGC 5813, NGC 2768, NGC 4340, NGC 4477, NGC
5746 and NGC 3631, which comprise 3 ellipticals, 3 S0s, 1 mid-type and 1
late-type spiral, respectively and can be found in the corresponding figures.

It is not unknown for some ellipticals to have colour gradients such that
their central regions are substantially redder than their outskirts
(e.g. Binney \& Merrifield, 1998). Here we investigate the emergence of this
effect in a pixel-by-pixel context. The luminosities of these galaxies are
compared with those of the rest of the sample using values quoted in Conselice
et al. (2000). All eight galaxies are bright but have luminosities typical of
the sample, in the range -19.17 $\leq$ M$_B$ $\leq$ -21.73.

A likely cause of the Red Hooks is dust reddening. In a study of 77 early-type
galaxies, Lauer et al. (2005) found that dust was present in the central
regions of about half of their sample. This is on a similar scale to that
which the central red pixels (CRPs) in the Frei sample are found, although a
much higher resolution was available in the HST/WFPC2 images. Nine of our
early-type galaxies are in both the Frei and Lauer et al. samples, three of
which are identified as having central red pixels. On the whole we find that
galaxies with no central reddening in our sample contain little or no dust,
whereas those identified as Red Hooks have visible dust structures obscuring
the central part of the galaxy, as identified from the Lauer et
al. study. This strongly suggests that dust is a major factor in causing the
Red Hook phenomenon. It is proposed by Lauer et al. that the dust features
that they have observed may outline gas that is falling into the central black
hole of the galaxy.

In general, those galaxies which appear to be very dusty in their images tend
to show a larger dispersion in (B - R) colour in their pCMDs, as opposed to
the neatness of the Red Hooks.  However, if these dust features exist in the
central regions of early-types, the orientation of such structures and the
amount of dust that is concentrated in the small central region may have a
different effect on the pCMD than large scale dust lanes. The observed
orientation of the dust rings and the amount by which they obscure the nucleus
of the galaxy could also influence the resulting pCMD. Although this is an
attractive idea, a more extensive and detailed study of this topic is required.

We also investigated whether the Red Hooks could be produced by AGN. Of the
eight galaxies that show this central reddening effect, six have known AGN
activity. Ho et al. (1997) studied a sample of more than 200 nearby galaxies
and classified their cores via spectroscopic measurements, including 56 of the
galaxies in our sample. Of the three elliptical galaxies found to have this
central reddening effect, NGC 4374 and NGC 5813 are both classified as L2 type
LINER cores whilst NGC 4125 is classified as a T2 transition object. According
to the arguments of Ho et al., transition objects are most naturally explained
as being ``normal'' LINERs whose spectra are diluted or contaminated by
neighbouring HII regions. Thus all three elliptical galaxies with a reddening
of their central pixels belong to the LINER class of galaxies.

Of the three S0s to show this effect, NGC 2768 is also classed as an L2 LINER
galaxy whilst NGC 4477 is classed as an S2 Seyfert galaxy. No information is
available for NGC 4340. NGC 5746, a mid-type spiral, is also classed as a T2
transition object whereas NGC 3631, a late-type spiral, has a HII core.

It is thus possible that interactions between the AGN and the interstellar gas in
the inner regions of these galaxies could affect the colours in these
regions. At least 34 out of the 69 galaxies in our sample are thought to have
some form of AGN activity, however, only six of them display the central
reddening effect.

A similar effect to the Red Hooks has been observed by e.g., Cantiello et
al. (2005), who looked at surface brightness fluctuations and colour gradients
of elliptical galaxies in the B and I bands. They also observed a steepening
in the (B-I) colour gradient in the centres of three elliptical galaxies. Such
colour gradients could provide a tool for investigating galaxy formation.

The fact that this steepening of the colour gradient is observed to
occur in only eight of the 69  galaxies analysed suggests that not all
galaxies have identical formation histories and/or properties, even within the
early-types. It is evident that this effect appears to be mostly associated
with early-type galaxies. This could imply that the central steepening of the
colour gradient is a phenomenon that emerges with time.

\subsubsection{NGC 4486 (M87) - AGN}
The pCMD of M87 (Figure 8) contains a branch of pixels that are bluer than
most other pixels at the same radius. These pixels are traced to a physical
region in the galaxy that appears as a bright spot, which is a manifestation
of the optical jet that originates from the central black hole of M87. This
jet is formed within a few tenths of a light year of the galaxy's core (Junor
et al. 1999).

\begin{figure*}
\centering
\includegraphics[angle=0, width=160mm, height=50mm]{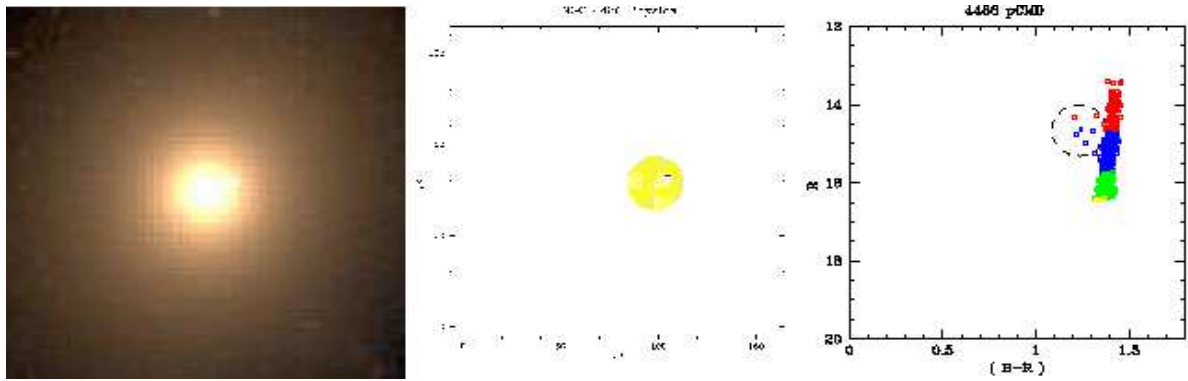}
\caption{A physical mapping of the blue branch of pixels in NGC 4486
  (M87). The image of NGC 4486 on the left is taken from the Frei et
  al. (1996) catalogue. The central image shows a physical plot of the galaxy
  (yellow pixels) with the blue branch of pixels overplotted in their
  corresponding colours, circled in the pCMD to the right.The few pixels
  that appear bluer than the main trend are shown to correspond to the optical
  jet of M87. }
\end{figure*}

\subsubsection{NGC 4826 - Extreme Dust}
 The pCMD of NGC 4826 (Figure 9) is unlike any other observed. The inner pixels of the
image appear to split into two branches clearly separated in both colour and
magnitude. The fainter, redder branch when physically mapped is revealed
to correspond to a region above the central bulge of the galaxy where there is
a prominent dust lane. Without this feature, the pCMD of NGC 4826 would be
typical of a mid-type galaxy.

\begin{figure*}
\centering
\includegraphics[angle=0, width=160mm, height=50mm]{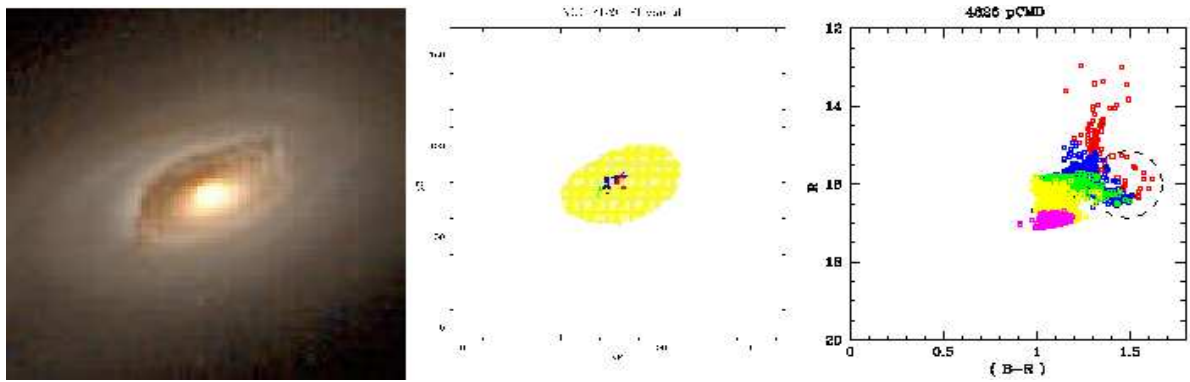}
\caption{A physical mapping of the red branch of pixels in NGC 4826.  The
  central image shows a physical plot of the galaxy (yellow pixels) with the
  red branch of pixels overplotted in their corresponding colours, circled in
  the pCMD to the right. The fainter, redder branch of pixels in the pCMD is
  shown to correspond to the prominent dust lane above the central bulge.}
\end{figure*}

\section{Quantifying pCMD Structure}
Through analysis of the information that can be gained from the pCMDs, we
construct new parameters formulated to pick out more fundamental properties of
galaxies and hence lead to a new, objective classification system. We  explore
a number of such methods, including the pixel blue-to-red light ratio and
colour spread, which we detail below.

\subsection{Average Colour Magnitude Diagram}
The average pixel colour and magnitude for each galaxy is calculated by considering
each pixel of the galaxy image. These average values for each galaxy are
plotted in Figure 10. We see a trend such that the early-types are brighter and
redder than the spirals. In turn, the mid-types tend to be redder and
brighter than the late-types. This points to a natural evolution along the
Hubble sequence and is not unexpected. 

This is a surface area weighted average as all pixels are summed
independently, without being given a weighting according to how bright they
are. Using a lower value for signal-to-noise has the effect of raising the
total R brightness value but the average (B-R) colour stays approximately the
same. Therefore, using a lower cut-off for surface brightness will have the
affect of shifting the relation up slightly, in approximately the same manner
for all galaxies. For a higher S/N, the colour will stay approximately the
same but the average R value will be lower.

\begin{figure}
\includegraphics[width=84mm]{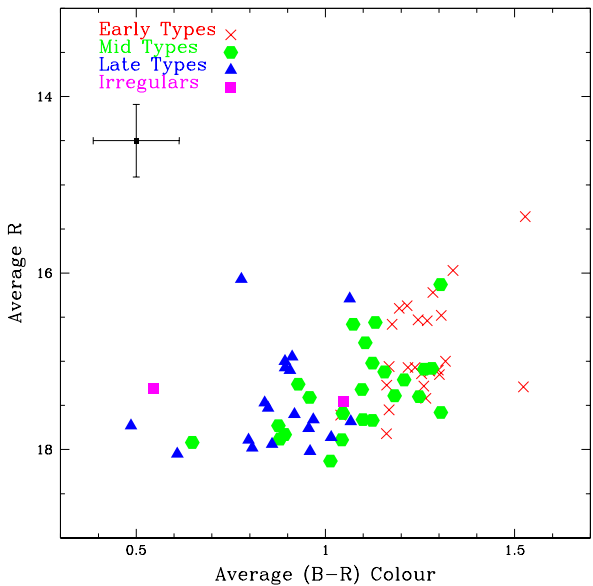}
\caption{The average colour magnitude diagram of the sample. Each point
  corresponds to the average colour and brightness of one galaxy, calculated
  on a pixel-by-pixel basis. The early-types are represented by diagonal
  crosses, mid-types by hexagons, late-type by triangles and the irregulars
  are shown by square points.}
\end{figure}

\subsection{Fitting the pCMDs}
We fit straight lines to the prime sequences of the pCMDs for most of our
galaxies, taking into account the error due to enhanced noise  at low
brightnesses. The pixels in each galaxy are binned into intervals of two
pixels in radius. The average colour and the average colour dispersion within
each of these intervals are then calculated and plotted.

These fits are a good representation of the behaviour we find in the pCMDs of
early-type galaxies but are obviously a poor representative of spiral and
irregular galaxies. The pCMDs of mid and late-type galaxies cannot be
characterised by such a simple functional form. The slopes tend to be steeper
for early-type galaxies but there is no clear pattern to distinguish
ellipticals from lenticulars. The irregular galaxies, NGC 3077 and NGC 4449
(Fig. 7), possess slopes that are opposite in sign to the early-types and
spirals due to the fact that they are bluest in their inner regions.

An example of each galaxy type is shown in Figure 11. Straight lines are fit
to all except the late-type spiral, NGC 3631, which clearly cannot be
described by such a fit. We note that the spread in colour of the S0 (NGC
3166) is larger than the elliptical (NGC 4472) and is more or less constant
throughout its radius. This confirms the observations noted in \S3.1.2.

\begin{figure*}
\centering
\includegraphics[angle=0, width=170mm, height=40mm]{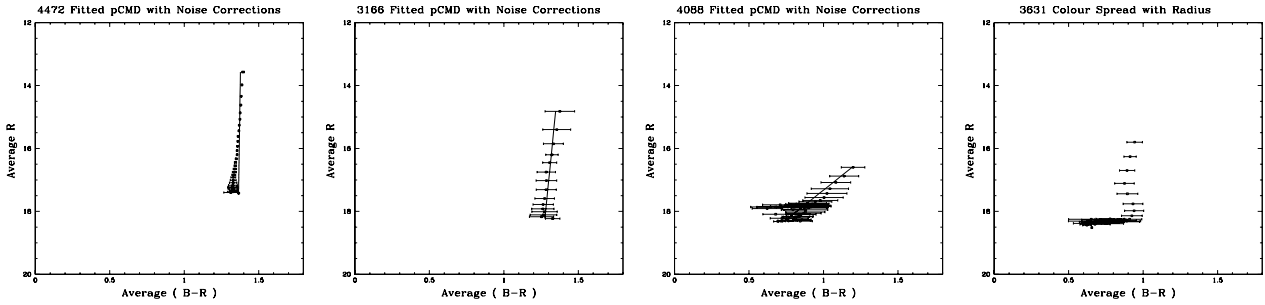}
\caption{Examples of the fits to the pCMDs. Above, from left to right, are NGC
  4472 (E), NGC 3166 (S0), NGC 4088 (mid-type) and NGC 3631 (late-type). The
  points represent the average (B-R) colour and the error bars represent the 1 sigma
  scatter in colour at various magnitudes. While a straight line fit works
  well for the early-types, it cannot be used to characterise the spirals.}
\end{figure*}

\subsection{Investigating the Average Colour Dispersion}
\subsubsection{Average Colour Dispersion Vs. Inclination Angle}
To test the relationship between dust content of a galaxy and spread in colour
in the pCMD, the average colour dispersion of the galaxy is compared with the
inclination (Fig. 12).  The average colour dispersion for each galaxy is
calculated by taking the mean of the spread in radial intervals as defined in
\S 4.2. The inclination angles are obtained from the online Frei catalogue
(1996). This test is limited to the mid and late-type spirals in the sample
so as to avoid any errors due to misclassification in the S0s.

In Fig. 12 the late-type spirals are plotted as blue triangles and the
mid-types as green hexagons. Despite the scatter in the plot, there is clearly
a positive trend between the average colour dispersion and inclination angle
of a galaxy. This implies that highly inclined galaxies are more likely to
have a large spread in colour in their pCMDs, and the greater the inclination,
the higher the internal spread in colour. This is evidence in favour of the
idea that a large scatter in colour in the pCMDs is partially due to the
influence of dust. However, we recognise that the colour dispersions are not
caused solely by dust, they are also influenced by differences in stellar
populations within the galaxy.

To test this idea, we compare the residuals of the inclination-average colour
dispersion fit (Fig. 12) with the average colour of the galaxies to check for
any systematic effects due to colour. We find no significant correlation,
which indicates that the scatter in the inclination Vs. colour relation is not
produced by differences in star formation histories.

\subsubsection{Average Colour Dispersion Vs. Average Colour}
We plot the average pixel (B-R) colour of each galaxy (as calculated in \S4.1)
against the average colour dispersion (Fig. 13). We  include the early-type
galaxies of the sample in this plot, represented by the red crosses. There is
a large scatter but there is a trend for redder galaxies to have a lower
average colour dispersion. There also appears to be a clear separation in the
average colour dispersion of early and late-type galaxies at around 0.05. This
could imply a fundamental difference between galaxies with star formation
present and those without. 

The late-type galaxy that has a particularly red colour and low dispersion for
its type is NGC 2775 (Fig. 6), which has a particularly faint disc compared to
the brightness of its central bulge. After signal-to-noise cuts only the inner
radius of 10 pixels from the galaxy centre remains, which corresponds to only
bulge light, giving the galaxy similar properties to that of an early-type in
this analysis. It is also possible that NGC 2775 could be  misclassified as a
mid-type.

\begin{figure}
\includegraphics[width=84mm]{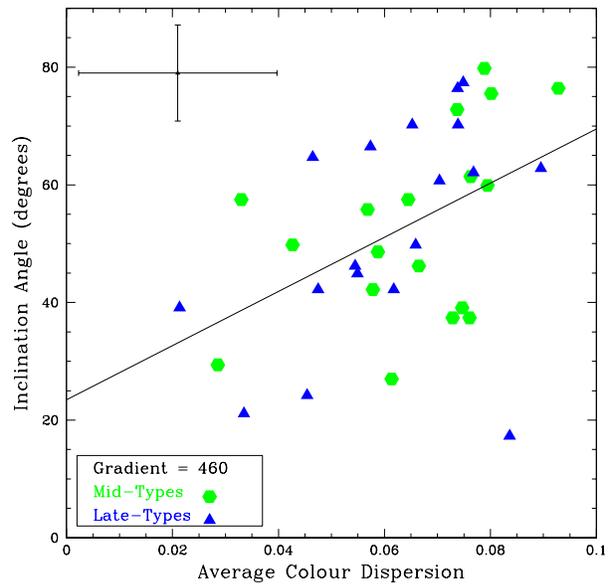}
\caption{The average dispersion in colour of each galaxy is plotted against
  inclination angle, as given in the Frei catalogue. Mid-type
  and late-type galaxies are plotted as green hexagons and blue triangles,
  respectively. A positive correlation can be distinguished but there is no
  discernable difference between the mid and late-types.}
\end{figure}

\begin{figure}
\includegraphics[width=84mm]{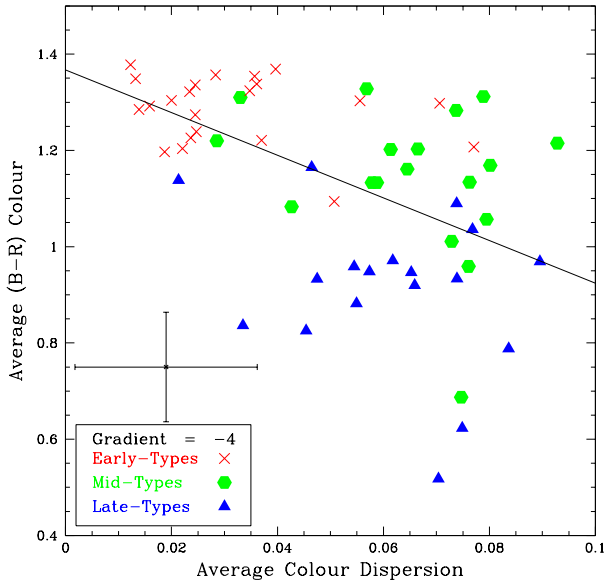}
\caption{The average (B-R) colour of each galaxy is plotted against the
  average dispersion in colour. Early-type galaxies are included in this plot
  and are plotted as red crosses. There is a trend for the redder galaxies to
  have a lower average colour dispersion. Late-type galaxies are distinguished
  from early and mid-types due to their blue colours.}
\end{figure}

\subsection{Blue-to-Red Light Ratio}
A central theme that emerges from the pCMDs is the fact that different
galaxies have different colour distributions and these are related to
morphological type. This is not a new idea, as it is well known that
early-type galaxies tend to be redder systems than late-type spirals, which
are hosts to star formation.  The analysis of bulge-to-disc ratios is another
useful measure of the properties of spiral galaxies, as found by Driver et
al. (2006). Analogous to this concept, we note that the ratio of total blue to
total red light in a galaxy could be a useful way of distinguishing between
galaxy types. We propose a new parameter to quantify this, the blue-to-red
light ratio.

A cut is made at (B - R) = 1, such that pixels with colours above that value
are defined as red light and those with colours below this value are defined
as blue light. In the Padova (1994) stellar population models with a Chabrier
IMF (Bruzual \& Charlot, 2003), which will be used in a further analysis later
on, (B - R) = 1 corresponds to an age of ~1 Gyr. This is the age at which
metallicity effects become important in colours, thus making (B - R) = 1 an
appropriate cut.  The blue-to-red light ratio for each galaxy is defined as
the difference between the total blue light and the total red light contained
in the pixels, divided by the total light. This gives a spatial measure of the
colours in the galaxy without being dependent on the intrinsic brightness of
the galaxy itself. We define the blue-to-red light ratio to be:

\begin{equation} \label{eq:BRrat}
\frac{N_B - N_R}{N_B + N_R}
= \frac{\sum_{n=1}^N p_n \arrowvert_{(B-R)<1} - \sum_{n=1}^N p_n \arrowvert_{(B-R)>1}} {\sum_{n=1}^N p_n}
\end{equation}

\noindent Where N$_B$ is the total number of pixels below the (B - R) = 1 cut,
N$_R$ is the total number of pixels above it, N is the total number of all
pixels and p represents each pixel in the galaxy image.

Figure 14 shows the average blue-to-red light ratio for each galaxy type: E,
S0, Spirals and Irregulars. A clear positive correlation can be seen, with
elliptical galaxies on average composed of mostly red light and the amount of
blue light contained in a galaxy increasing for later Hubble types. The
blue-to-red light ratio can be used to sort galaxies into early or late-type
and this could be an appropriate measure when looking at irregular structures
at high redshift.

\begin{figure}
\includegraphics[width=84mm]{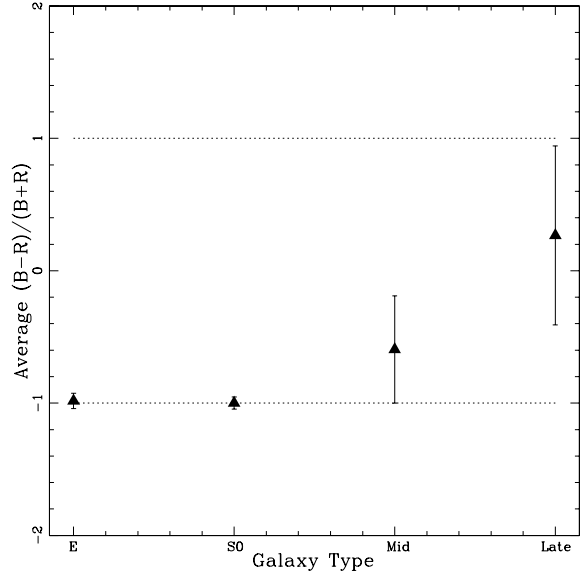}
\caption{The average blue to red light ratio for each galaxy type.}
\end{figure}

\subsection{Galaxy Colour Distributions}

\begin{figure}
\includegraphics[width=84mm]{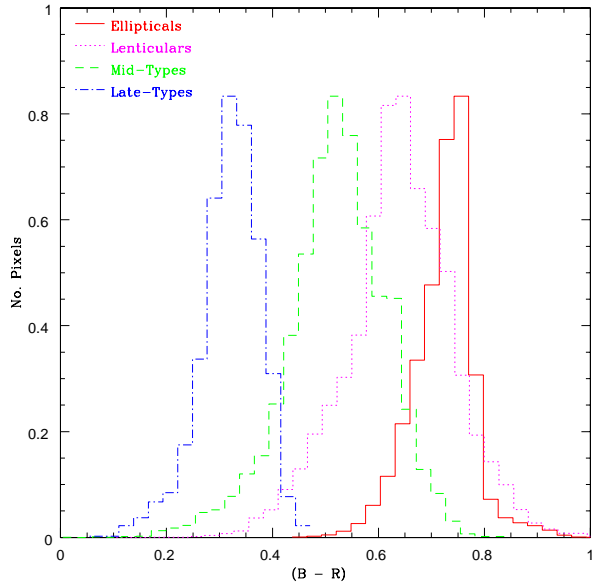}
\caption{The total colour distributions for pixels, sorted into galaxy
types.}
\end{figure}

Figure 15 shows the cumulative distribution of colour for all galaxy types,
taking each pixel into account. There is almost no overlap in colour between
elliptical galaxies and late-type spirals, whereas the mid-type and S0 peaks
lie between these with significant overlap. There appears to be a natural
progression in colour from late-type (blue) to early-type (red) along the
Hubble sequence.

It is interesting to compare this plot to the work of Driver et al. (2006) who
examined bulge-disc compositions of early-type spirals. Driver et al. see a
clear separation between the bulge and disc of early-type spirals, with bulges
occupying their ``red compact peak'' and discs their ``blue diffuse
peak''. However, the S0 colour distribution in Fig. 15 significantly overlaps
with the elliptical and mid-type colour distributions and hardly at all with
the blue, late-type distribution. This implies that the bulges of lenticular
galaxies are as red (or redder, perhaps due to dust) than some pixels of
elliptical galaxies but there is no clear blue peak of pixels relating to the
S0 disc component.

We also do not see clear and distinct bulge and disc colour bimodality in a
pixel analysis of (B - R) colour distribution for mid or late-types. There
appears, rather, to be a steady sequence in colour across each galaxy type.

\section{Comparison to Stellar Population Models}

We use the stellar population models of Bruzual \& Charlot (2003), hereafter
BC03, to obtain very rough relative age estimates of the pixels in each
galaxy. We use the Padova (1994) model with a Chabrier IMF. The available
metallicities in these models are sub-solar, solar and approximately twice
solar. None of the galaxies in our sample are likely to be sub-solar, on
average, as they are all massive and luminous. It is also impossible to
calculate metallicities on a pixel-by-pixel basis with only two photometric
bands. Therefore, for this preliminary analysis and for comparison purposes,
solar metallicity is used for all pixels. We do, however, recognise that
internal galaxy metallicity gradients are ignored in this analysis.

Figure 16 shows the relationship between age and (B-R) colour from BC03
models used in the analysis. Overplotted on these points are our parametric
fits between these two quantities. Colours above (B - R) = 0.75 are fit to a
log profile and colours equal to and below this value are fit to a straight
line in order to reduce the degeneracy in low colour values. The form
of these fits are shown below.

\begin{displaymath} 
(B - R) = \left\{\begin{array}{cr}
 0.485\log({\rm age/Gyr}) + 1.08, & (B - R) > 0.75 \\
 1.566({\rm age/Gyr}) + 0.033, & (B - R) \leq 0.75 \\
  \end{array}\right.
\end{displaymath}

Each pixel is sorted by morphological type and the equivalent ages are
calculated. The distribution of ages of each galaxy type are discussed in the
next section.  

The models are not well fit for the region around 5 Gyr. However, the
relation is monotonic and we are interested in relative ages only. Therefore, the
systematic offset introduced in this region does not present a problem for our
purpose in this analysis.

\begin{figure}
\includegraphics[width=84mm]{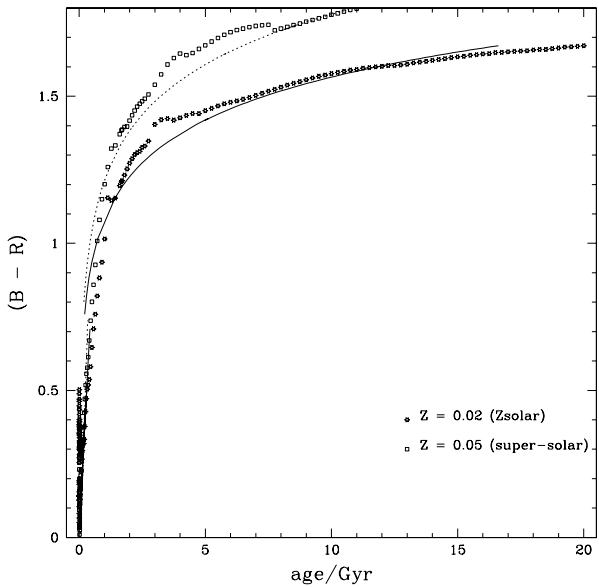}
\caption{Fitting the BC03 models, with a Chabrier IMF. Colours above (B
  - R) = 0.75 were fit to a log(x) profile whereas colours below this were fit
  as straight lines.}
\end{figure}

\subsection{Age Distributions of Pixels}

Figure 17 shows the distribution of ages for all pixels in our sample, sorted
via morphological type. We see a clear separation in age for each galaxy
type. We recognise that the difficulties of estimating ages with only two
photometric bands means that the absolute ages stated here should not be taken
literally. The comparison of the pixel colours with the BC03 models yield ages
which are too low to comply with current age estimates, for example, of
stellar populations in globular clusters (Sharina et al. 2006, Strader et
al. 2005, Krauss \& Chaboyer 2003). Rather, these results are more useful for
looking at the relative ages of pixels in each galaxy type. One reason these
fits give such young ages is due to the large degeneracy between age and
colour after 6 Gyr.

The ellipticals appear to be the oldest, with the majority of their pixels
having been formed significantly before those of all the other types. The S0
population appears to be the next oldest with a distribution similar to that
of the mid-types but shifted in age. The distribution of the late-types, with
fitted ages less than 1 Gyr, are so blue they fall into the very steepest
region of the BC03 models (see Fig. 16).

However, there are inherent biases in this pixel analysis. Younger stars tend
to be bluer and much brighter than older, redder stars and hence will have
more influence on the measured colour of a pixel. Some pixels will also
naturally contain more stars than others and will contribute more flux purely
due to numbers. The former bias is impossible to fix but adjustments can be
made for the latter by altering Fig. 17 to account for the amount of stellar
mass in each pixel. This is carried out in the next section.

\begin{figure}
\includegraphics[width=84mm]{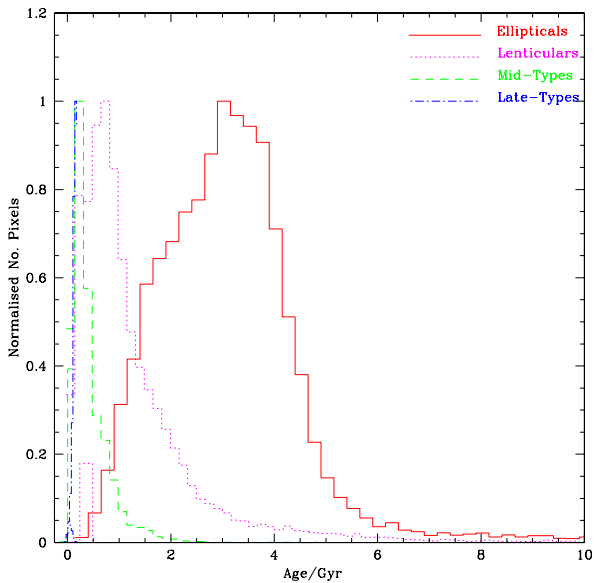}
\caption{The age distribution of the pixels of all galaxies, sorted into
  morphological type. These ages were produced using the BC03 models assuming
  solar metallicity.}
\end{figure}

\subsection{Mass-Weighted Age Distribution}
The stellar mass in each pixel of the galaxy images is calculated using the
Bell \& deJong (2001) relation for the mass-to-light ratio of galaxies as a
function of  (B - R) colour, as listed in equation (\ref{eq:BdeJ}).

\begin{equation} \label{eq:BdeJ}
{\rm log_{10}}\Big(\frac{M}{L}\Big) = -1.224 +1.251(B-R)
\end{equation}

\noindent The total mass in each age bin is calculated and these values are normalised
for each galaxy type. The normalised mass values are plotted against the
mid-point of each age bin and the results are shown in Fig. 18. The effect of
this mass-weighting is to increase the ages of some pixels. The peak age of
the ellipticals, when weighted by mass, increases by approximately 0.5 Gyr,
whereas the peak of the mid-types has now been increased to coincide with that
of the S0s. All late-type pixels still fall at the lowest ages due to their
very blue colours.

Interestingly, the effect of mass-weighting the S0s is to spread the age
distribution. There is a trend of steady increase in total stellar mass up to
$\sim 1.5$ Gyr and then a rapid decline up to the present epoch. This is in
contrast to the ellipticals, which seem to have a limited range in age, with
little or no mass formation at the present epoch. This tail in the S0s
suggests that many of their pixels are older than those of the
ellipticals. However, this is not likely to be real, rather it results from
some of the S0 pixels having very red (B-R) colours due to the presence of
dust. This tail of old S0 pixels can be seen in Figure 15. The pixels that
have these extreme red colours in the S0s are also some of the brightest,
hence they become more pronounced in the mass-weighted age diagram.

\begin{figure}
\includegraphics[width=84mm]{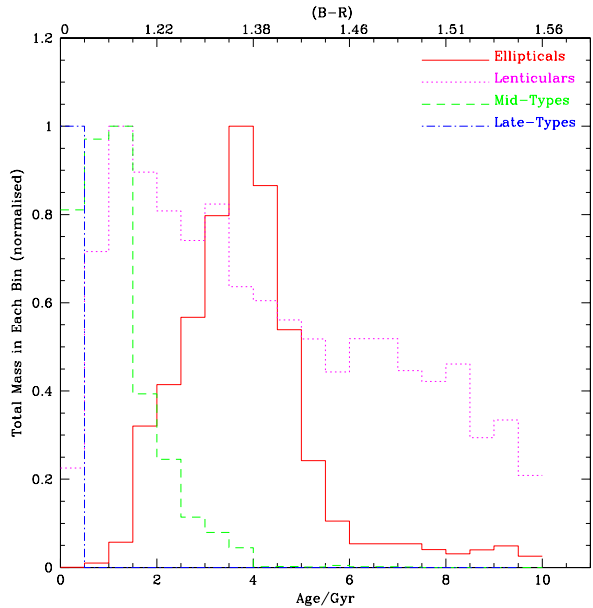}
\caption{Mass-weighted age distribution of the pixels of all galaxies, sorted
  into morphological type. Stellar masses of each pixel were calculated using
  the Bell \& DeJong (2001) model for calculating mass-to-light ratios for B and R
  band luminosities. The stellar masses of each pixel were then binned according to
  age and summed.}
\end{figure}

\subsection{Stellar Population Maps}

Pixel-by-pixel analysis allows us to investigate galaxy properties that cannot
be probed using integrated light alone. Using the masses of each pixel,
calculated in \S 5.2, we can construct maps of the stellar mass distribution
and M/L ratio distribution throughout each galaxy. Figure 19 illustrates
comparisons between galaxy images, stellar mass disributions and mass-to-light
ratio maps for five galaxies representative of each morphological type. The
scale of both the stellar mass and M/L maps is set such that the brighter
pixels represent those with larger masses and higher M/L ratios. The masses
and M/L maps are calculated with the negative pixel values removed. This is so
that low S/N pixels in the images can still be mapped.

\begin{figure*}
\centering
\includegraphics[angle=0, width=120mm, height=200mm]{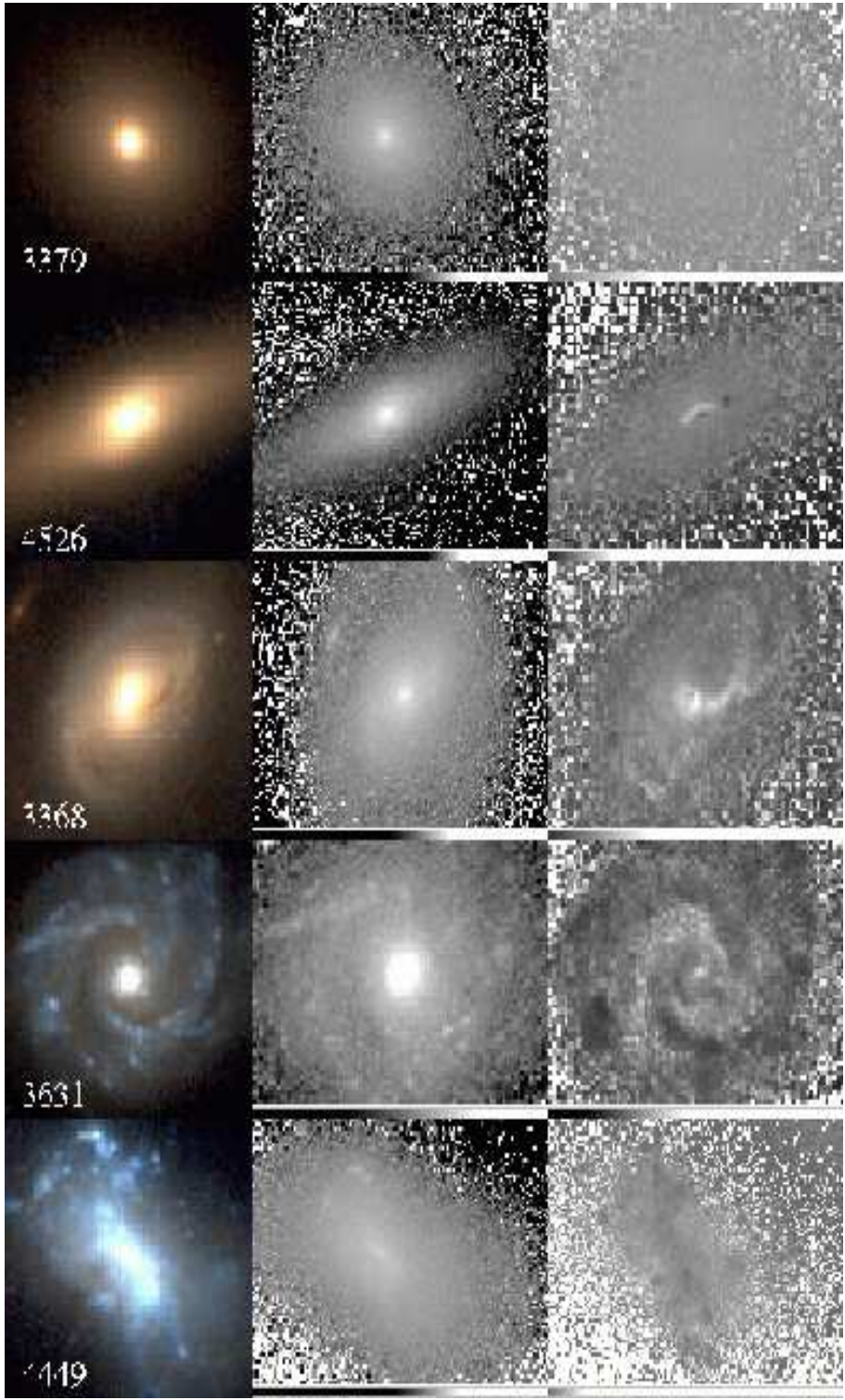}
\caption{A comparison between the light, stellar mass and mass-to-light
  distributions of galaxies across the Hubble sequence, arranged from left to
  right. The brighter pixels indicate the greatest masses/highest mass-to-light ratios.
  The galaxies shown (from the top) are NGC3379 (E), NGC4526 (S0),
  NGC3368 (mid-type), NGC3631 (late-type) and NGC4449 (irregular).}
\end{figure*}

\subsubsection{Stellar Mass Maps}
In general for these example galaxies the stellar mass distributions follow
the light in the images (Figure 19). The most massive pixels lie close to the
centres of all galaxies. Spiral structure can still be identified in NGC 3368
although the mass distribution is smoother than that of the light, suggesting
that concentrations of light do not always correspond to concentrations of
mass. Star formation will dominate the galaxy image, as discussed in \S 5.1,
but the mass appears to be more evenly spread throughout the disc. This is
confirmed in the mass map of NGC 3631 where the spiral arms are not so
prevalent, instead the mass of the disc is approximately constant throughout
the disc, with the exception of a few small knots. The same is true of the
irregular galaxy NGC 4449, whose mass distribution appears much smoother than
the light in the optical image.

\subsubsection{Mass-to-Light Ratio Maps}
The M/L ratio maps show more variation in their distributions than the stellar
mass maps. We note that all pixel in the M/L ratio maps of NGC 3379 (E), NGC 4526
(S0) and NGC 3368 (mid-type) have values of M/L $>$ 1. However, for NGC 3631 (late-type) 
and NGC 4449 (irregular) all the pixels in the maps have values of M/L between 0 and 1.
In all maps, the scale indicates that the brightest pixels have the highest values 
of M/L.

The M/L ratio is approximately constant in the early-type galaxies. Although the
few highest valued pixels in the S0 lie in the central regions of the
galaxy a central concentration has disappeared altogether in the
elliptical. This implies that, while there is more mass than light throughout
NGC 3379, the bright central pixels evident in the Frei image and the mass map
are not only very massive, they are also very bright. 

NGC 3368 yields an interesting M/L map (Fig. 19), displaying a bias in M/L
ratio in one spiral arm. This is explained by the presence of dust on that
side of the galaxy, as seen in the light image. This dustier, redder section
of the galaxy will appear older and will appear to have a boosted M/L ratio
for that reason.

Star formation dominates most of the pixels in the M/L maps of NGC 3631 and
NGC 4449 such that the brightest pixels in the Frei image tend to have
low M/L values. The bulge almost disappears in the map of NGC 3631 implying
that the stars in this region are still relatively young although the spiral
structure that was diminished in the mass map reappears in the M/L map.

\section{Application to the UDF}

For the pCMD technique to be truly useful as an analysis and classification
tool, it must be transferable between photometric bands and redshifts. We
test this condition by applying the method to the Hubble Ultra Deep Field
(HUDF) in the i(775nm) and z(850nm) bands. One early-type galaxy and two
spirals are chosen. The pCMDs are constructed from these images using the
method described in \S 2. The resulting pCMDs are shown in Figure 20.

The early-type galaxy does not appear any redder in (i-z) than the spirals,
although its radial profile is more uniform and its gradient steeper. The
spirals do not seem to show as much structure in their pCMDs as would be
expected from the findings of \S 3. This could be due to the smaller
difference in wavelength between i and z compared to the diffence between the
B and R bands in the Lowell sample.It could also be due to the different
underlying stellar populations in the high-z when compared with the local
galaxies. The effects of ``beam dilution'' can be discounted in this instance
as the typical scale subtended by a pixel in the HUDF galaxies chosen is of
the order of that in the local sample. This is partially due to the much
greater resolution of the ACS than the nearby galaxy images taken from the
ground.

This exercise demonstrates that a pixel-by-pixel analysis can be performed
over a wide range of redshifts and photometric bands.

\begin{figure*}
\centering
\includegraphics[angle=0, width=160mm, height=50mm]{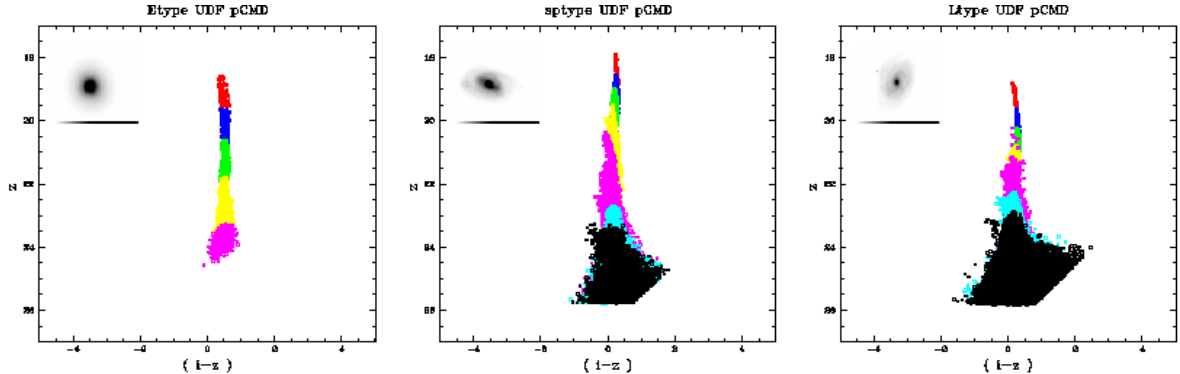}
\caption{The pCMDs of 3 galaxies chosen from the UDF in the i and z bands.}
\end{figure*}

\section{Discussion and Conclusions}

\subsection{pCMDs}

Pixel colour magnitude diagrams provide a novel way of looking at the stellar
populations and structures of galaxies. We find that the pCMDs of each
galaxy type are distinct and trends have been identified.

Elliptical pCMDs tend to have a small spread in colour throughout their radii
and have nearly uniform, red colours, which we call the ``prime
sequence''. The lenticular galaxies in the sample are similar to the
elliptical pCMDs but with a greater internal colour spread. The mid and
late-type pCMDs tend to be bluer than the E and S0s and have a greater colour
gradient throughout their radius, such that they are bluer toward the edges of
the galaxies and redder in the central regions. The mid and late-type pCMDs
display more varied structures. Trends within these types are identified.

We find that ``inverse-L'' shape pCMDs correspond to face-on spiral
galaxies and the distinct red and blue regions of the pCMDs are caused by the
bulge and disc components of the galaxies. Dusty and high inclination spirals
produce pCMDs with a large spread in colour throughout their radii and this
could also explain the greater color spread of the lenticular
pCMDs. Individual features of two galaxies are identified through their
pCMDs. These are the optical jet in NGC4486 and a prominent dust lane in
NGC4826. Red Hook features are identified in 6 early-type galaxies. These
are attributed to dust in the central regions of the galaxies.

\subsection{Pixel-by-Pixel Analysis}

Pixel-by-pixel analysis is a useful way of gaining information from galaxy
images and reveals information that analysis of integrated light does not. In
this paper, we explore methods that can be used as part of an automated
classification system for galaxies using pixels and define some new
parameters, such as the blue-to-red light ratio.

We look at the internal colour distributions of galaxies in a new way. The
colour distributions of all pixels are organised by morphological type and we
find that these alone show no bimodality in colour within one type. We confirm
that there does appear to be a steady progression in average pixel colour
along the Hubble sequence, with late-types exhibiting the bluest colours and
ellipticals the reddest, as expected.

We compare our galaxy pixel colours to the Bruzual \& Charlot (2003)
stellar population models and use this to calculate ages and
mass-weighted ages for each pixel, organised by type. We find that elliptical
galaxies have the oldest and late-types the youngest pixels.

The pixel-by-pixel technique allows the stellar mass distribution and M/L
ratio in galaxies to be mapped. We find that the mass tends to follow the
light in the cases of the five example galaxies tested. This is not the case
with the M/L maps, which show trends related to both star formation and dust extinction.
Pixel-by-pixel mass maps is a more fundamental way of looking at
galaxies and it will be very interesting to apply to galaxies at high redshift.
 
\subsection{Application to High-z Galaxies}
Finally, we create pCMDs for three galaxies in the HUDF in the i(775) and z(850)
bands. This proves that the pCMD analysis can transcend different data
sets, different bands and can be applied to galaxies at high redshift.

\subsection{Future Work}
There is much more that can be done using a pixel approach to analysing galaxy
image data. This work presents new questions about the morphological evolution
of galaxies that need to be answered. For example, further investigation of
the central regions of nearby early-type galaxies is needed to explore the
origins of the Red Hooks.

An in-depth analysis of individual galaxies would be very interesting using a
pixel approach, as an extension to this and previous work by Bothun (1986) and
Eskridge et al. (2003). For example, looking at HST data of nearby galaxies,
such as M51, could allow for a pCMD that reveals further features that
integrated light analyses cannot.

The pixel approach could be further exploited by formulating new structural
parameters of galaxies. This would be an extension of the CAS system
introduced by Conselice, Bershady and Jangren (Conselice et al., 2000,
Bershady et al., 2000) where parameters are calculated pixel-by-pixel instead
of through the use of light profiles, etc. In combination with the further
development of stellar mass maps and their application to high redshift
galaxies, a pixel approach will provide a more fundamental way of
investigating and classifying galaxies.

\section*{Acknowledgments}

We would like to thank Reynier Peletier for useful communications about the
nature of the centres of early-type galaxies. We would also like to thank the
referee for their very helpful comments and suggestions. MML acknowledges
postgraduate funding from the Particle Physics and Astronomy Research Council
(UK).

\end{document}